\documentclass[aps,prl,amsmath,amssymb,notitlepage,twocolumn,superscriptaddress]{revtex4-1}
\usepackage{amsmath,epsfig,amssymb}
\usepackage{bbm}
\usepackage{bm}
\usepackage{times}
\usepackage{color}
\usepackage{amsthm}
\usepackage{color}
\usepackage[caption=false]{subfig}
\usepackage{multirow}
\usepackage[normalem]{ulem}
\usepackage{qcircuit}
\usepackage{braket}
\usepackage{amsfonts,amssymb,dsfont}
\usepackage{graphicx}
\usepackage{soul}
\usepackage{hyperref}
\usepackage{inputenc}

\usepackage[acronym,shortcuts]{glossaries}

\hypersetup{
    colorlinks=true,       
    linkcolor=red,          
  citecolor=magenta,        
    filecolor=magenta,      
    urlcolor=cyan,           
    runcolor=cyan
}

\usepackage[capitalise,compress]{cleveref}
\crefname{section}{Sec.}{Secs.}
\Crefname{section}{Section}{Sections}

\newcommand{\todo}[1]{}
\renewcommand{\todo}[1]{{\color{red} TODO: {#1}}}
\newcommand{\question}[1]{}
\renewcommand{\question}[1]{{\color{red} QUESTION: {#1}}}

\newacronym{SchWARMA}{SchWARMA}{Schr\"{o}dinger Wave Autoregressive Moving
Average}
\newacronym{ARMA}{ARMA}{Autoregressive Moving Average}
\newacronym{FTTPS}{FTTPS}{Fixed Total Time Pulse Sequences}
\newacronym{RFTTPS}{RFTTPS}{Robust FTTPS}
\newacronym{NISQ}{NISQ}{Noisy Intermediate-Scale Quantum}
\newacronym{SDR}{SDR}{Software Defined Radio}
\newacronym{AWG}{AWG}{Arbitrary Waveform Generator}
\newacronym{QNS}{QNS}{Quantum Noise Spectroscopy}

\begin{document} 

\glsdisablehyper

%
\title{Universal Dephasing Noise Injection via Schrodinger Wave Autoregressive Moving Average Models}
\author{Andrew Murphy}
\affiliation{
Johns Hopkins University Applied Physics Laboratory,
Laurel, MD, 20723, USA
}
\author{Jacob Epstein}
\affiliation{
Johns Hopkins University Applied Physics Laboratory,
Laurel, MD, 20723, USA
}
\author{Gregory Quiroz}
\affiliation{
Johns Hopkins University Applied Physics Laboratory,
Laurel, MD, 20723, USA
}
\author{Kevin Schultz}
\affiliation{
Johns Hopkins University Applied Physics Laboratory,
Laurel, MD, 20723, USA
}
\author{Lina Tewala}
\affiliation{
Johns Hopkins University Applied Physics Laboratory,
Laurel, MD, 20723, USA
}
\affiliation{Thomas C. Jenkins Department of Biophysics,
Johns Hopkins University,
Baltimore, MD, 21218, USA
}
\author{Kyle McElroy}
\affiliation{
Johns Hopkins University Applied Physics Laboratory,
Laurel, MD, 20723, USA
}
\author{Colin Trout}
\affiliation{
Johns Hopkins University Applied Physics Laboratory,
Laurel, MD, 20723, USA
}
\author{Brian Tien-Street}
\affiliation{
Johns Hopkins University Applied Physics Laboratory,
Laurel, MD, 20723, USA
}
\author{Joan A. Hoffmann}
\affiliation{
Johns Hopkins University Applied Physics Laboratory,
Laurel, MD, 20723, USA
}
\author{B. D. Clader}
\affiliation{
Johns Hopkins University Applied Physics Laboratory,
Laurel, MD, 20723, USA
}
\author{Junling Long}
\affiliation{
National Institute of Standards and Technology, Boulder, Colorado 80305, USA
}
\author{David P. Pappas}
\affiliation{
National Institute of Standards and Technology, Boulder, Colorado 80305, USA
}
\author{Timothy M. Sweeney}
\affiliation{
Johns Hopkins University Applied Physics Laboratory,
Laurel, MD, 20723, USA
}

\date{\today}

\maketitle
\textbf{
We present and validate a novel method for noise injection of arbitrary spectra in quantum circuits that can be applied to any system capable of executing arbitrary single qubit rotations, including cloud-based quantum processors. As the consequences of temporally-correlated noise on the performance of quantum algorithms are not well understood, the capability to engineer and inject such noise in quantum systems is paramount. To date, noise injection capabilities have been limited and highly platform specific, requiring low-level access to control hardware. We experimentally validate our universal method by comparing to a direct hardware-based noise-injection scheme, using a combination of quantum noise spectroscopy and classical signal analysis to show that the two approaches agree. These results showcase a highly versatile method for noise injection that can be utilized by theoretical and experimental researchers to verify, evaluate, and improve quantum characterization protocols and quantum algorithms for sensing and computing.}

As qubit technology advances into the \ac{NISQ} era, spurred by quantum
algorithms that promise advantage over classical alternatives, quantum hardware
is progressively able to support circuits that are increasingly complex in
circuit depth and system size. While numerous physical technologies and
architectures are being developed, each system suffers from limitations and
sensitivities due to environmental and systematic noise sources. Ultimately,
these noise sources lead to computational errors that destroy the efficacy of a
quantum computation. For this reason, a significant amount of research has
focused on characterizing \cite{romach2015:qns, Frey2017:qns, chan2018:qns,
sung2019:qns, lupke2019:qns, blumekohout2011:gst, knill2008:rb, magesan2011:rb}
and addressing (i.e., avoiding \cite{lidar2014:dfs-rev}, suppressing
\cite{viola1998:dd,viola1999:dd,zanardi1999:dd,viola2013:dd-rev}, or correcting
\cite{shor1995:qec,steane1996:qec,aliferis2006:qec,terhal2015:2015,nielsen2011:qc} noise in quantum systems. These
efforts have been complemented by evaluations of quantum algorithmic resilience
to specific noise sources \cite{chuang1995:qc-err, knill1998:qc-err,
Knill1998:qc-err2, colless2018:qc-err, mcclean2017:qc-err}.

\begin{figure*}[t]
\centering
\includegraphics[width=\textwidth]{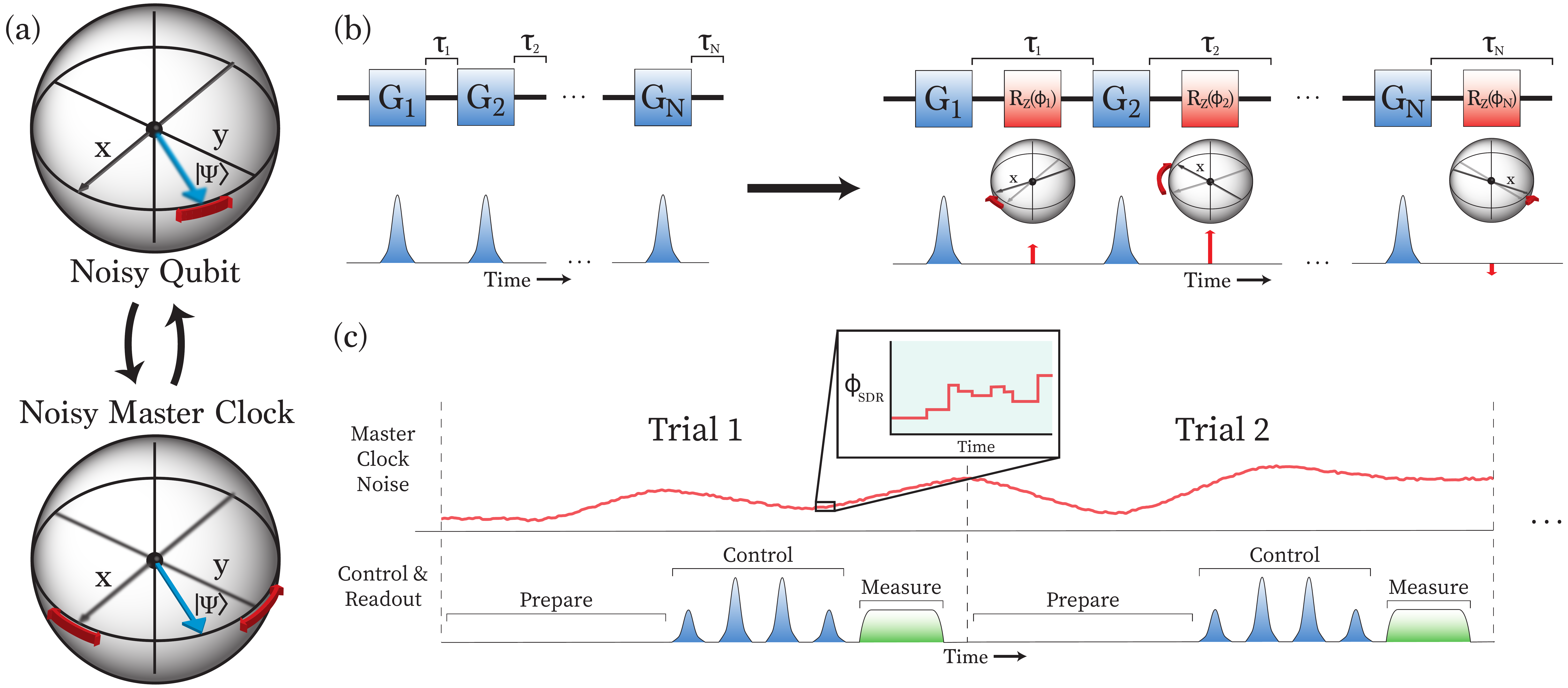}
\caption{ Implementation of SchWARMA. (a) Phase noise on a
    qubit under ideal measurement conditions (upper Bloch sphere) is simulated by imparting controlled errors
    on the experimental clock (i.e., the reference frame) with no intentional additional noise on the
    qubit (lower Bloch sphere). (b) Gate Injection. SchWARMA models an ideal quantum circuit in a noisy environment by interleaving error gates between each control pulse. We implement these error gates as instantaneous rotations of the qubit's reference frame. (c) SDR-based noise injection. In the SDR setup, the SDR signal is mixed with control pulses to impart errors in the control signal. The phase of the SDR signal is updated discretely, every 100 ns, but the signal itself runs continuously over all shots in an experiment. Unlike gate-injection experiments, each shot in an SDR-based noise injection experiment will have a unique phase trajectory of the control signal.}
\label{fig:main-diagram}
\end{figure*}

Engineered noise provides a necessary capability for
investigating the effects of noise in a controlled environment. Beyond
providing a means for validation and verification of system characterization
protocols, engineered noise allows one to examine the susceptibility and
robustness of noise characterization and mitigation protocols to particular
noise types \cite{mavadia2018:gst-err}. Furthermore, when sufficiently
versatile, engineered noise can provide key insight into inherent noise
resilience and vulnerabilities of quantum algorithms to targeted noise
sources. In these ways, engineered noise enables the development of robust
system characterization techniques, noise mitigation protocols, and error resilient
quantum algorithms, all of which can be considered necessary for achieving
fault-tolerant quantum computation.
To date, engineered noise has been achieved in a few experimental
platforms, but a standard method for generating well-defined noise of arbitrary
spectra has yet to be identified. Furthermore, methods for injecting engineered noise have
been highly platform specific \cite{sung2019:qns, lupke2019:qns}, with
particular focus on lab-based experimental systems of superconducting transmon
qubits \cite{sung2019:qns, lupke2019:qns} or trapped ions
\cite{soare2014:eng-noise}. These noise injection protocols require
either a specific device design \cite{sung2019:qns, lupke2019:qns}, or control of
specific in-house hardware platforms \cite{soare2014:eng-noise}.

In this article, we present a platform-agnostic technique for injecting
engineered dephasing noise of arbitrary spectra based on the \ac{SchWARMA} model
\cite{schwarma}. A generalization of the \ac{ARMA} models used in classical
signal processing, \ac{SchWARMA} models have been previously used to
numerically simulate spatio-temporally correlated noise processes in
multi-qubit systems.
%
%
Exploiting the framework of \ac{SchWARMA}-based numerical simulation, we show that the concept of inserting \ac{SchWARMA}-designed error gates into an ideal quantum circuit can be leveraged for noise injection. This is a key advantage of SchWARMA over other noise injection techniques,
making it implementable in any situation where access to arbitrary single qubit
rotation gates is provided, including cloud devices such as the IBM Quantum
Experience (IBMQE) \cite{ibmqe}.
Focusing in particular on phase noise, we show that \ac{SchWARMA} can be utilized  to introduce temporally
correlated dephasing errors in \textit{real} physical systems, and that the \ac{SchWARMA}
approach has excellent predictive power
as well. While this study centers around stationary Gaussian phase noise, we
emphasize that the \ac{SchWARMA}  approach is far-reaching and enables
injection of control amplitude noise, as well as non-stationary, non-Gaussian,
and spatiotemporally correlated noise.

To validate the \ac{SchWARMA} model as both a statistical model for
describing temporally-correlated noise and as a generative model for mimicking
phase-noise on a qubit, we develop and implement two
\ac{SchWARMA}-motivated noise injection approaches: (1) a gate-based approach --
faithful to the original \ac{SchWARMA} concept -- that inserts correlated phase
errors into an ideal circuit, and (2) a low-cost
hardware implementation that uses a \ac{SDR} to impart errors directly on the master clock. Approach (2) is less universal and not platform-agnostic, requiring direct access to hardware for implementation. However, unlike (1), method (2) allows for direct measurement of the injected noise spectrum. For this reason, we present method (1) as a proof-of-principle for arbitrary phase noise spectra injection using \ac{SchWARMA} and present method (2) as a validator for method (1). The IBMQE serves as a testbed for approach (1), where hardware access is purposefully restricted to preset gates to convey the universality of SchWARMA noise injection on systems with limited access. In addition, an in-house transmon-based superconducting qubit system, which we will refer to as the Applied Physics Laboratory (APL) system, is employed as an evaluation testbed that allows for both gate-based and direct access scenarios to be examined and compared.

%
%
Using the qubit as a probe, the injected noise spectra is reconstructed via \ac{QNS} \cite{alvarez2011:qns, romach2015:qns,Frey2017:qns, chan2018:qns, sung2019:qns, lupke2019:qns}. We find strong agreement between the reconstructed and injected noise spectra for both gate- and SDR-based approaches. The SDR approach is further validated by direct measurement of its output with a signal analyzer, displaying agreement with the desired spectrum.
Additionally, we show that forward simulations of the corresponding
\ac{SchWARMA} models have strong correlation with the experimental data,
closing the loop on the validation of the \ac{SchWARMA} approach for modeling
temporally correlated dephasing noise.\\

\noindent\textbf{Results}\\
\noindent\textbf{Dephasing Noise Injection.}
The dephasing between a qubit and a master clock can be engineered through either dynamic control of a qubit's phase relative to the master clock or through the dynamic control of the phase of the master clock relative to the qubit.  The former case  is represented by the upper panel of Fig.~\ref{fig:main-diagram}(a) where the reference frame defined by the local oscillator is stationary, and the qubit state wanders about the z-axis.  Conversely, in the latter case, represented by the lower panel of Fig.~\ref{fig:main-diagram}(a), the qubit state is stationary while the reference frame wanders~\cite{ball2016:clock}. We artificially induce
correlated phase noise using this latter method, through dynamical control over the master clock on a single qubit system. The
result is an effective qubit response that mimics that of a qubit subject to a temporally correlated classical noise
environment.
The phase
modulation of the master clock $\phi(t)=\phi_\text{ctrl}(t) +
\phi_\text{noise}(t)$ is composed of the typical (potentially faulty) control
phase $\phi_\text{ctrl}(t)$, with an additional $\phi_\text{noise}(t)$
representing the desired injected noise process. Next, we discuss how the
\ac{SchWARMA} formalism can be used to used to construct $\phi_{\text{noise}}(t)$.\\

\begin{figure}[t]
\centering
\includegraphics[width=\linewidth]{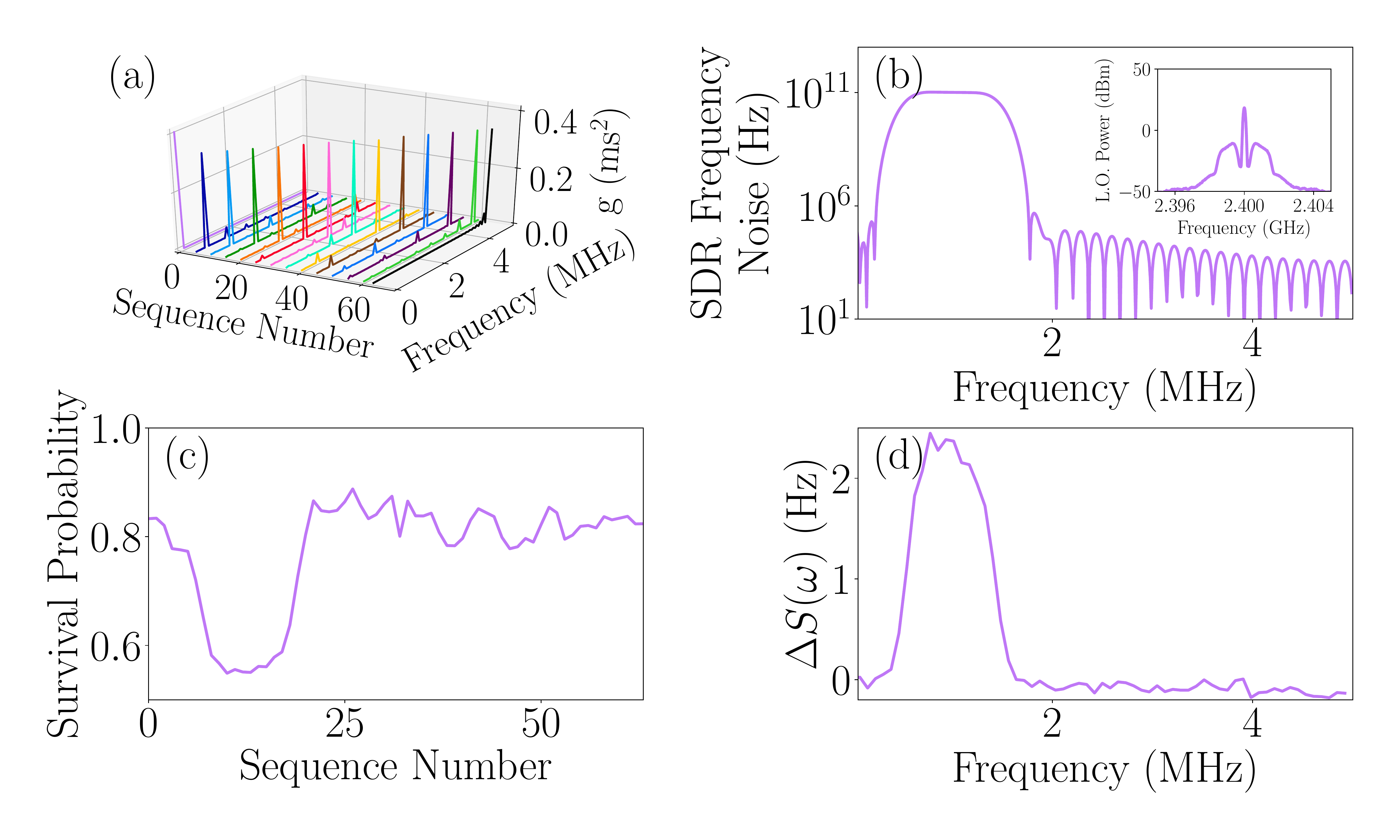}
\caption{SchWARMA based noise injection and reconstruction. (a) The filter functions of the fixed total-time pulse sequences have sharp and unique dependencies on noise of different frequencies. (b) Noise of a desired spectrum, in this example bandpass, is generated with either the gate-based or SDR-based injection method. Inset: A measurement of the frequency spectrum of the 2.4 GHz noisy carrier signal out of the SDR. This signal is mixed with an ideal control signal and upconverted to the qubit frequency. (c) The survival probabilities for each sequence can be analyzed to reconstruct the noise influencing the system (d). }
\label{fig:recon}
\end{figure}

\noindent\textbf{Noise Injection via \ac{SchWARMA}.} 
%
\ac{ARMA} models are widely used in the field of time series
analysis to model time correlations in data. \ac{SchWARMA} provides a natural
path for generalization of these classic techniques from signal processing to quantum
information. While \ac{SchWARMA} has been previously utilized as an approach
for numerically simulating classical temporally correlated noise in quantum circuits, its
utility can be readily extended to noise injection and model prediction. 

The gate-based noise injection protocol follows a similar recipe to the numerical approach demonstrated in \cite{schwarma}, where the noise is injected via error
gates interleaved in a quantum circuit, as is demonstrated in Fig.~\ref{fig:main-diagram}(b). In the case of single qubit dephasing,
a circuit with noise injection can be expressed as
\begin{equation} \label{eq:schwarma} 
    U(\bm{\phi}) =  R_z(\phi_N) G_N  R_z(\phi_{N-1})   G_{N-1} \cdots
    R_z(\phi_1) G_1
\end{equation} 
where $G_j$ denotes a single qubit operation in the noiseless circuit and the
error gate $R_z(\phi_j)$ represents a $z$-rotation with time-correlated
$\phi_j$. Upon specifying properties of the noise process, e.g., the mean and
power spectral density, a \ac{SchWARMA} model is used to generate a trajectory
of temporally correlated phases $\bm{\phi}=\{\phi_1, \ldots, \phi_N\}$.
Averaging an observable or fidelity metric over many realizations of
$\bm{\phi}$ results in the desired dephasing behavior.

This form of \ac{SchWARMA}-based noise injection is equivalent to control
master clock phase noise injection. In superconducting qubits, an $R_z(\phi)$
gate is implemented via a virtual frame change that simply updates the control
phase (Fig.~\ref{fig:main-diagram}a). In this sense, $\bm{\phi}$ represents a discretized implementation of
$\phi_{\text{noise}}(t)$. When the sampling time of
$\phi_{\text{noise}}(t)$ is the fundamental gate time, $t_G$,  $\bm{\phi}$ and
$\phi_{\text{noise}}(t)$ are identical. Note that all gates $G_j$ do not need
to be equivalent in duration, but rather just integer multiples of $t_G$. For
example, if gate $G_j$ requires timing $n t_G$, the composition of $n$ error
gates are appended to the circuit following $G_j$. 


Below, we also introduce an \ac{SDR}-based \ac{SchWARMA} noise injection
technique that is distinct from the gate-based approach in that it occurs
simultaneously and asynchronously with the control. Requiring lower level
control access, the \ac{SDR} approach involves injecting phase noise directly
into the control lines. As a result, the sampling time of the noise
is \emph{not} necessarily restricted by the gate time $t_G$. However, even in
the case where the sampling time is set by $t_G$, the injected
noise and gates are inherently asynchronous; thus, phase updates may frequently
occur during pulses. Because the engineered noise in the \ac{SDR}-based noise injection technique is produced independently of the control pulses, this method allows a more direct comparison between the engineered and measured noise via a classical signal analyzer, a method that cannot be leveraged in the gate-based approach. In addition to these distinct features, we elaborate on
other facets of the \ac{SDR} approach in the
discussion section.\\


\noindent\textbf{Noise Injection Validation.}
The \ac{SchWARMA} noise injection
protocol is validated via \ac{QNS}.
We utilize a set of deterministically generated \ac{FTTPS} designed to probe the noise spectrum across a frequency band of interest. Each sequence $S_k$ is comprised of $\pi$-pulses that produce a unique filter function $g_k$ that is well-concentrated in frequency space [Fig.~\ref{fig:recon}(a)]. The FTTPS probe the noise spectrum by altering the number of pulses within a fixed total sequence time. This protocol is distinct from the Carr-Purcell-Meiboom-Gill (CPMG) based \ac{QNS} protocol \cite{alvarez2011:qns} that probes the noise spectrum by varying the interpulse delay (and total sequence time) using a fixed number of $\pi$-pulses, and it is distinct from other fixed total time sequences found from a random search~\cite{sung2019:qns}. Further details on \ac{FTTPS} are provided in the supplement.

Dephasing noise [Fig.~\ref{fig:recon}(b)] is injected into a single qubit system via the \ac{SDR} or gate-based \ac{SchWARMA} procedure, thereby corrupting the FTTPS. Survival probabilities are then measured for the FTTPS [Fig.~\ref{fig:recon}(c)], where the survival probability for each probe sequence exhibits a distinct (and sharp) dependence on noise of specific frequencies. The resulting survival probabilities are used to construct an estimated noise spectrum [Fig.~\ref{fig:recon}(d)]. Note that we define the survival probability as the probability of the qubit being in a desired state at the end of a probe FTTPS.

Below, we
demonstrate that \ac{QNS} reconstructions agree with the various SchWARMA generated and injected noise spectra.
In addition, we show
that \ac{SchWARMA} can be used to achieve a detailed prediction of the
\ac{FTTPS} survival probabilities.
In order to separate injected noise from the native background, the native
noise spectrum $S_{\text{nat}}(\omega)$ is measured using the procedure
described above and subtracted from the injected noise spectrum reconstructions $S_{\text{inj}}(\omega)$ to obtain $\Delta S(\omega)
=S_{\text{inj}}(\omega) -S_{\text{nat}}(\omega)$. \\


\noindent\textbf{Gate-Based Noise Injection.}
Gate-based noise injection provides a means for examining quantum algorithms in the presence of engineered noise when hardware access is limited to particular gate operations. Here, we demonstrate gate-based noise injection via \ac{SchWARMA}. The IBMQE serves as a restricted access testbed, where we intentionally utilize the circuit-level interface to limit circuit operations to the standard IBM gate set \cite{mckay2017:gates}, demonstrating the availability of the approach to the general user community. Phase noise is introduced through $U_1(\lambda)$ phase gates that are interleaved with the gates that comprise the sequences that define the probe \ac{FTTPS}. The time resolution of
the noise is set by the $X$ gate timing, given by the $U_3(\theta,\phi,\lambda)$
gate time. The $U_3$ gate implements generic unitary operations using three
configurable frame changes interleaved with $R_x(\pi/2)$ and $R_x(-\pi/2)$ rotations. The
frame changes are used to set the phases $\theta, \phi, \lambda$. 

\begin{figure}[t]
  \centering
  \includegraphics[width=\linewidth]{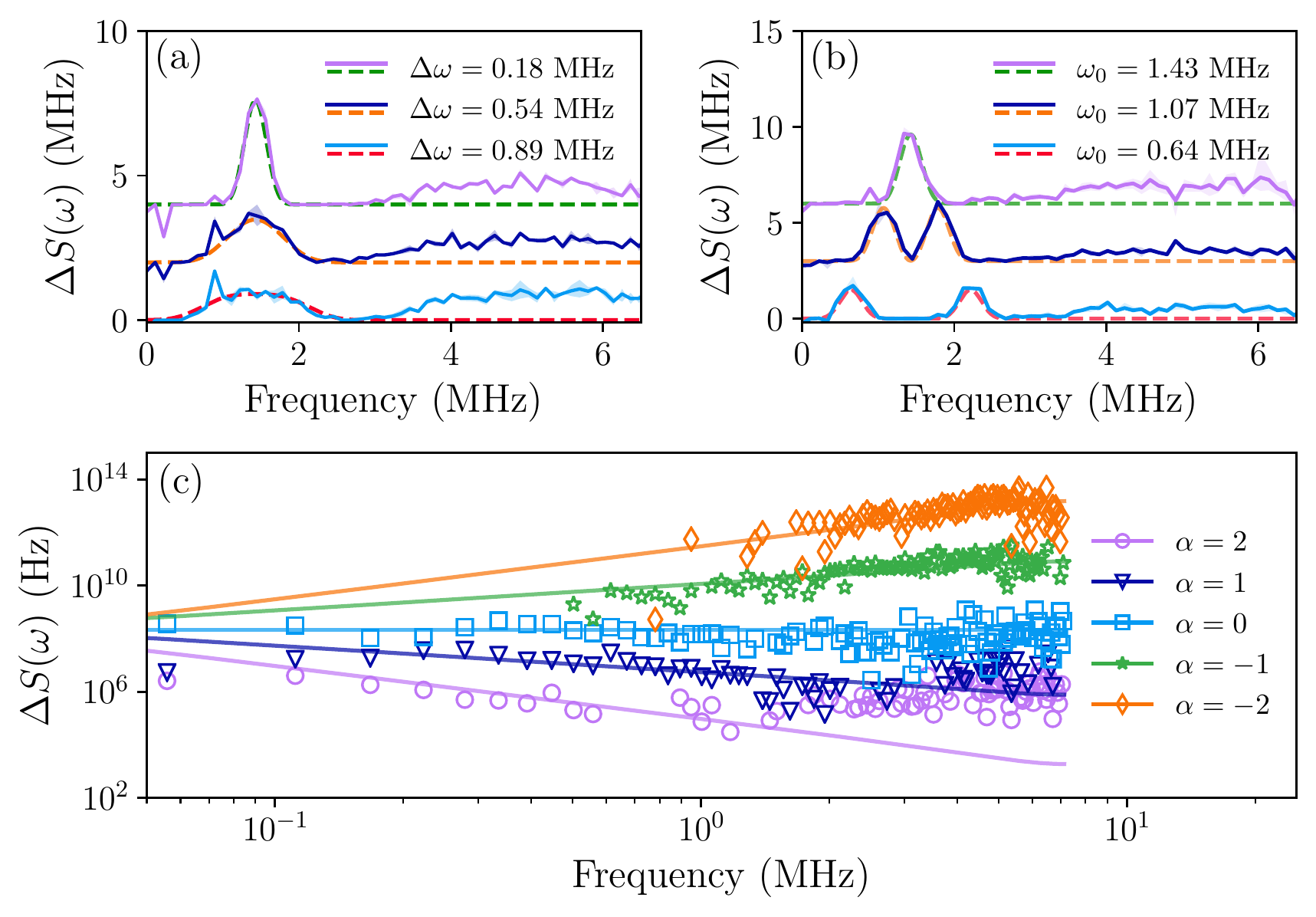}
  \caption{\ac{SchWARMA} gate-based noise injection on the IBMQE. Noise reconstructions are shown for (a) bandpass, (b) double bandpass, and (c) $1/f^\alpha$ noise for different bandwidths, center frequencies, and spectral decays. Ideal spectra (dotted lines in panels (a) and (b) and solid lines in panel (c)) are found to be in good agreement with estimated spectra (solid lines in panels (a) and (b) and symbols in panel (c)). The data shown here are compiled over five calibration cycles, with solid lines denoting bootstrapped medians and shaded regions specifying $95\%$ confidence intervals in panels (a) and (b). Note that the confidence intervals are small for a majority of the estimated spectra. Data has been offset for clarity.}
  \label{fig:ibm-inj}
\end{figure}

Employing the Vigo processor for our experiment, where $t_{G}\approx 70$ ns, we
prepare a single qubit in the equal superposition state, apply the \ac{FTTPS}
with injected phase noise, apply a gate that would return the state to $\ket{0}$ in the absence of noise, and
measure. The experiment is performed over 50 independent realizations of \ac{SchWARMA}
trajectories for each probe sequence, where statistics for each experiment are
collected over 1000 shots. The complete noise injection and noise spectrum
estimation experiment is conducted over five calibration cycles and
bootstrapped to mitigate spectral anomalies
solely due to variations in hardware characteristics across calibration cycles, thereby improving spectral estimates.
Refer to the supplement for further information regarding optimized
experimental practices and procedures.


Through the \ac{SchWARMA} gate-based injected noise protocol, we find good
agreement between the desired and reconstructed noise spectra. In
Figure~\ref{fig:ibm-inj}, we show the results for three distinct classes of noise spectra: (a) a single bandpass spectrum of varying width, (b)
a multi-bandpass spectrum with varying center frequencies, and (c)
$1/f^{\alpha}$ noise. We consider single bandpass bandwidths of
$\Delta\omega=0.18,0.54,0.89$ MHz, with a center frequency of $\omega_0=1.43$
MHz. 
We explore the versatility of \ac{SchWARMA} by separating the injected noise
into two bandpass regions centered at $\omega_0=1.07$ MHz and $\omega_1=1.79$
MHz, and subsequently, $\omega_0=0.64$ MHz and $\omega_1=2.21$ MHz, while
keeping the bandwidth of 0.18~MHz constant for each region.
Lastly, we consider injected $1/f^\alpha$-type noise with $\alpha=-2,-1,0,1,2$.
Such frequency dependence is commonly found in superconducting
~\cite{burnett2014:fnoise, burnett2019:fnoise, muller2015:fnoise,
bylander2011:fnoise, yan2013:fnoise} and semiconductor~\cite{basset2014:fnoise,
chan2018:qns, struck2020:fnoise} qubit systems.

The resulting  spectral density estimates yield spectral features, such as peak
height, bandwidth, and center frequency, that are found to be in good agreement
with the desired noise profiles. Additional spectral features observed at high
frequency have been determined to be artificial, resulting from increased gate
error as higher frequency regimes are probed with sequences containing a
greater number of pulses. The manifestation of gate error in dephasing noise
spectra has been previously noted and mitigated using spin-locking (SL)
techniques~\cite{yan2016:qns}. IBM's quantum control
toolbox~\cite{thomas2020:pulses} provides a means for exploring SL-based methods, as well as pulse-error compensating sequences for QNS. We focus on the latter and create a set of \ac{RFTTPS} composed of calibrated, true $R_x(\pi)$ and $R_x(-\pi)$ pulses.  It has long been known in the NMR and dynamical decoupling communties that pulse phase adjustments can aid in the suppression of pulse imperfection errors (e.g., over/under-rotation errors) for pulse-based error mitigation schemes ~\cite{vold1973:comp, levitt1981:comp, taylor2007:thesis, quiroz2013:gadd}. By constructing \ac{RFTTPS} composed of pulses of alternating phases (i.e., $R_x(\pi)$ and $R_x(-\pi)$), we effectively reduce the effect of pulse imperfections on the estimated spectra. \ac{RFTTPS} are employed on the APL system in the subsequent section, and an additional comparison between \ac{FTTPS} and \ac{RFTTPS} using on the IBMQE is included in the supplement. The use of \ac{RFTTPS} along with additional simulations (see supplement) support the conclusion that the observed high-frequency noise is primarily an artifact of gate error.\\

\begin{figure}[t]
  \centering
  \includegraphics[width=\linewidth]{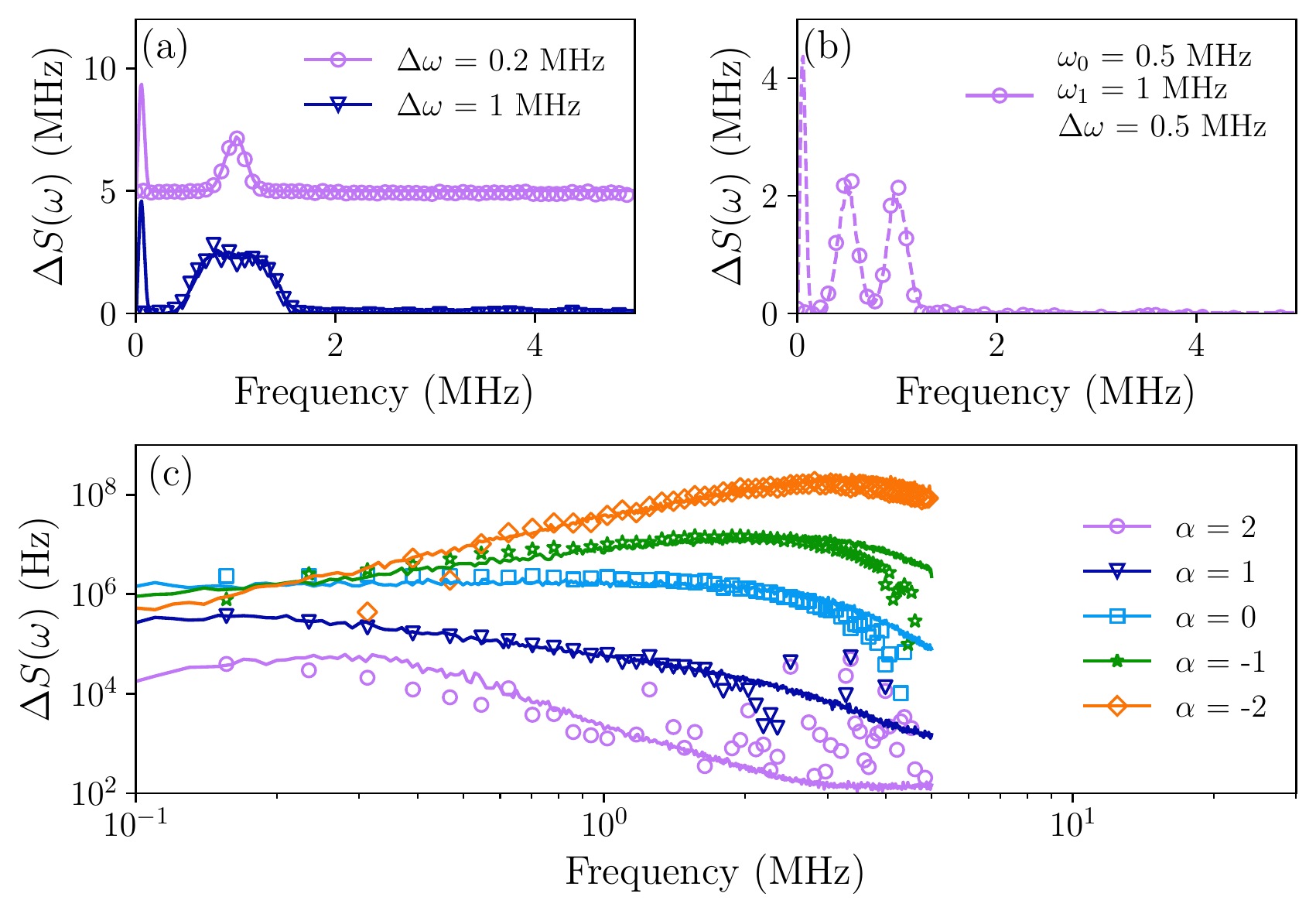}
  \caption{Verification of \ac{SDR}-based \ac{SchWARMA} noise injection. Noise
    reconstructions are shown for (a) bandpass, (b) double bandpass, and (c)
    $1/f^\alpha$ noise for different bandwidths, center frequencies and spectral decays. The data shown are reconstructed from the average survival probability of 10,000 unique shots per sequence, each with a unique phase trajectory. Points represent noise reconstructions, and solid lines are measurements performed on the generated noise by a signal analyzer. In figures (a) and (c) plots are offset for clarity.}
  \label{fig:sdr}
\end{figure}

\noindent\textbf{Verification with \ac{SDR}.} 
%
Our SDR-based approach to noise injection relies on direct hardware access. This approach is not platform agnostic and is presented as a means to validate SchWARMA gate-based noise injection by directly comparing results against a control line signal with the synthesized noise spectrum.
In order to meet the requirements of lower-level access, we turn to the APL system 
to perform this comparison.

 An \ac{SDR} is configured to
implement a \ac{SchWARMA} model and produce $\phi_{\text{noise}}(t)$.  
The control signal sums the desired control phase with the phase-noise spectrum
output of the \ac{SDR}, acting as a faulty control signal with a drifting
master clock. The \ac{SDR} outputs a continuous 2.4 GHz carrier signal with pseudo-random phase changes that are updated every 100 ns. While the phase changes may be random, one advantage of \ac{SchWARMA} is that temporal correlations are permitted, meaning that the pseudo-random phase changes are biased by previously applied phase updates that mimic the dephasing error gates in the gate-based approach. 
In software, these fluctuations are shaped from
white-noise into an engineered, \ac{SchWARMA}-generated noise spectrum. We
measure the \ac{SDR} output on a signal analyzer with a 90 kHz bandwidth as a
ground-truth comparison for spectral reconstructions  (Fig.~\ref{fig:recon}(b) inset).



After confirmation via signal analyzer, 
we perform a series of \ac{RFTTPS} qubit experiments in the APL system with the \ac{SDR} as the master clock and the clock phase controlled by various \ac{SchWARMA} models. The APL system includes a fixed-frequency transmon qubit held at 20 mK at the mixing chamber stage of a dilution refrigerator. The qubit ($\approx 5.4$ GHz) is coupled to a readout resonator ($\approx 7$ GHz) and the state of the qubit is determined by dispersive measurement techniques. See the supplement for further detail about \ac{SDR} injection and the APL experimental setup.  Experiments using the \ac{SDR} approach are performed over 10,000 unique shots, and the $R_x(\pm\pi)$ pulses are calibrated to have $t_G=100$ ns for both \ac{SDR} injection and gate-based injection in the APL experimental setup. The fixed total time of noise injection experiments in the APL setup is 12.8 $\mu$s, well within the measured coherence times of the qubit ($T_1 \approx 47$ $\mu$s and $T_2 \approx 58$ $\mu$s). Before measurement, we apply a $R_x(\pi/2)$ gate which would excite an ideally behaved qubit in the absence of noise to the $\ket{1}$ state. Using the same \ac{QNS} analysis applied in the gate-based  approach we reconstruct the injected noise spectra.
As is the case with the gate-based approach, the \ac{SDR} injection approach
can be used to produce a variety of phase-noise spectra. In
Fig.~\ref{fig:sdr}, we show the results for (a) a single bandpass spectrum of varying width and noise power, (b) a multi-bandpass spectrum, and (c) $1/f^\alpha$ noise. We consider single bandpass bandwidths of
$\Delta\omega=0.2,1$~MHz with a center frequency $\omega_0 = 1 $~MHz.  The double bandpass noise spectrum is engineered to have center frequencies
$\omega_0=0.5$~MHz and $\omega_1=1$ MHz, with noise bandwidths
$\Delta\omega=0.5$ MHz. Lastly, we consider $1/f^\alpha$-type noise with  $\alpha=-2,-1,0,1,2$, as we did with gate-based experiments.

Reconstructed spectra are compared against the measured frequency-noise spectrum of the
\ac{SDR} output signal (measured on a signal analyzer) for validation. To achieve fits, the measured frequency spectra of the \ac{SDR} output signal are each multiplied by a scaling factor, constant over experimental setup, to account for attenuation of the signal between room temperature and the qubit. Overall, the reconstructed noise
spectra match the injected spectra well. The measured bandwidths, peak heights, and center frequencies of the injected bandpass and double bandpass spectra are in good agreement with the injected noise spectra. The behavior of $1/f^\alpha$ spectra match the
expected power dependencies best within approximately $0.1$~MHz $\leq \omega_0\leq
2$~MHz  (Fig.~\ref{fig:sdr} (c)). At frequencies above 2 MHz, the innate roll off in the \ac{SDR} output
distorts the injected noise, leading to inaccuracies in the reconstructed noise
spectra. Despite these distortions, the overall shape of the reconstructed
spectra remains in good agreement with the expected noise spectra measured at the output of the \ac{SDR} (solid lines).

\begin{figure}[t]
  \centering
  \includegraphics[width=\linewidth]{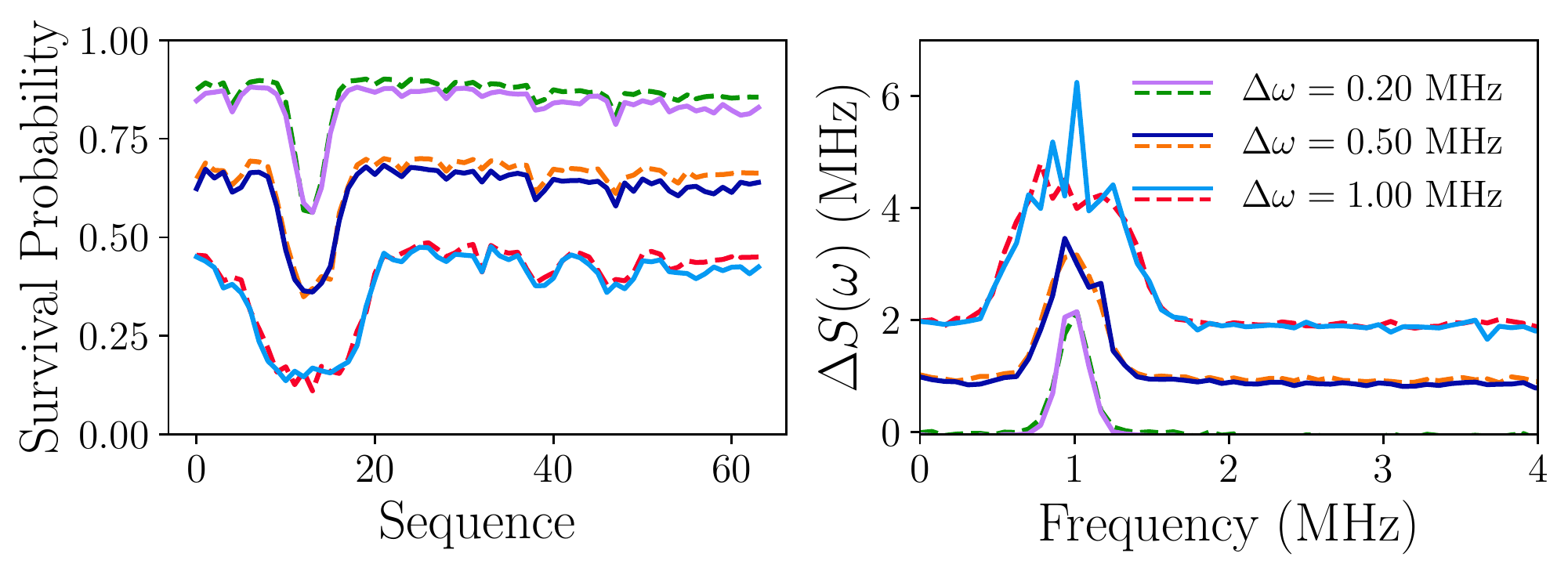}
  \caption{Comparison between gate-based (solid lines) and \ac{SDR}-based
    (dashed lines) \ac{SchWARMA} noise injection for bandpass dephasing
    noise with three different bandwidths. Good agreement between the approaches is found for all spectra considered; thus, validating SchWARMA gate-based noise injection. Data for $\Delta\omega=0.5,1.0$ MHz are vertically offset by 0.2 for survival probabilities (left) and 1 MHz for spectrum reconstructions (right).}
  \label{fig:gate-sdr}
\end{figure}

\begin{figure}[t]
  \centering
  \includegraphics[width=\linewidth]{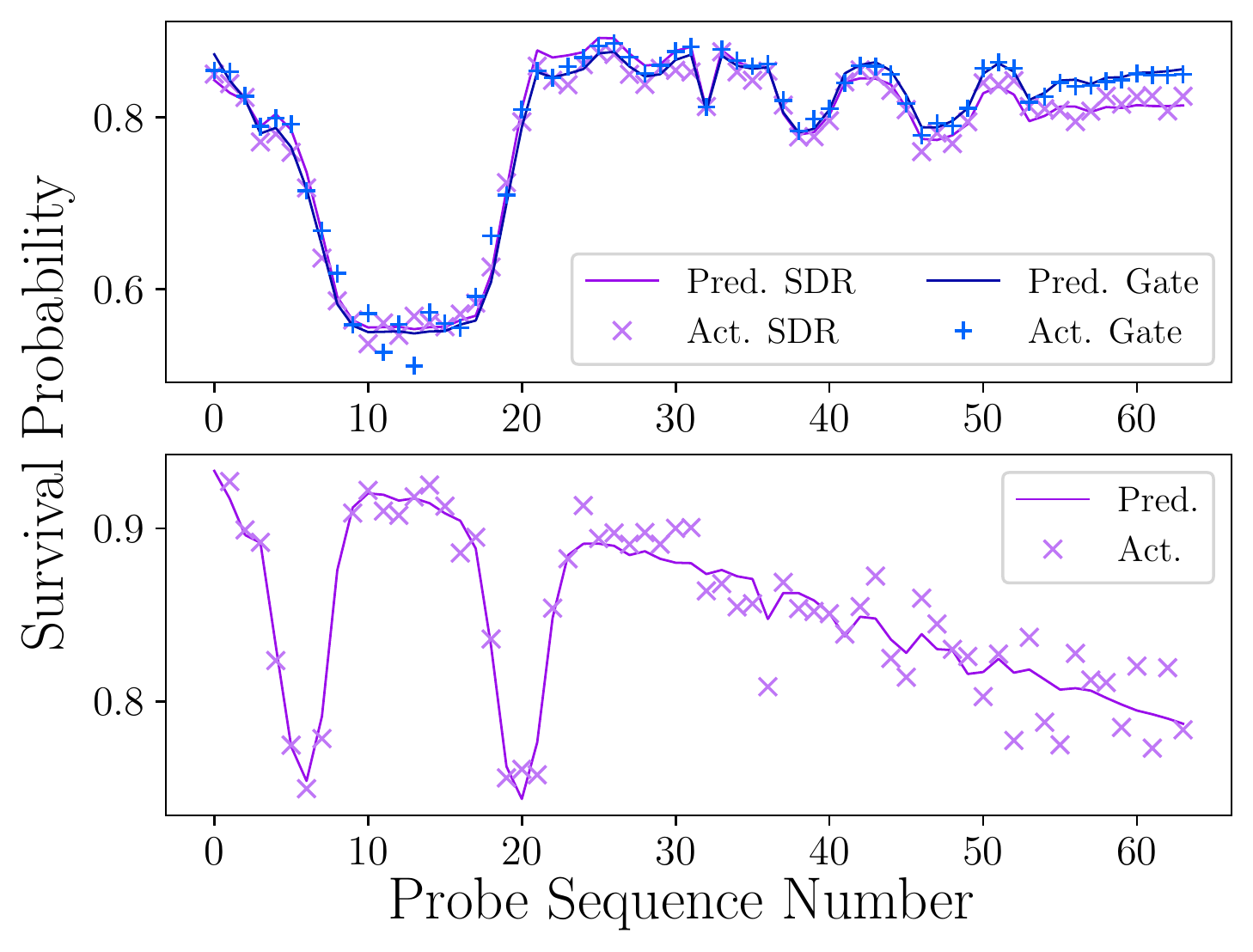}
  \caption{Comparison of \ac{SchWARMA} predicted to actual
    experimentally measured survival probabilities using \ac{FTTPS}.  
    \textbf{Top:}
    \ac{SDR}- and gate-based noise injection of a wide bandpass spectrum on the APL
    experimental platform.
    \textbf{Bottom:} Gate-based injection of a double bandpass spectrum on the IBMQE.
  }
  \label{fig:schwarma_preds}
\end{figure}

The combination of the signal analyzer output and the reconstructed spectra
provides confidence that the \ac{SDR}-based approach injects phase noise in the
desired manner.  We can transfer this confidence to the gate-based approach by
comparing reconstructions of gate-based noise injection of identical spectra on
the same device.
%
%
Fig.~\ref{fig:gate-sdr} shows reconstructions of gate-based \ac{SchWARMA}
injected noise plotted on top of nominally identical noise injected using the
\ac{SDR} experimental approach. Gate-based injection experiments performed on the APL experimental setup are measured over 200 unique SchWARMA  trajectories, with 1000 shots per trajectory. We reiterate that both measurements in Fig.~\ref{fig:gate-sdr} are
performed on the same APL experimental setup. These results show excellent agreement
between the two methods, with the primary discrepancy between the approaches
due to finite sampling effects on the measurements and number of \ac{SchWARMA}
trajectories used in the gate-based approach. Further comparisons between the gate-based and SDR approaches are included in the supplement.\\
\noindent\textbf{\ac{SchWARMA} Model Prediction.}
Much like classical \ac{ARMA}, a \ac{SchWARMA} model's associated power
spectrum can be used to model real physical systems and predict in closed form
their average response to given stimuli (here pulse sequences). This
effectively closes the loop between the generation and injection processes.
Here, using the known injected \ac{SchWARMA} model, we fit a few ancillary
parameters to experimentally generated \ac{FTTPS} data.
Fig.~\ref{fig:schwarma_preds} shows the results of this fitting process for
sample experiments performed via SDR and gate-based noise injection
(details of this process are found in the following paragraphs). 
The top panel shows SDR and gate-based data for the 1~MHz bandpass experiments whose reconstructions are shown in Figs.~\ref{fig:sdr} and \ref{fig:gate-sdr}, and the bottom panel displays gate-based data from the double bandpass experiment with $\omega_0$=1.07~MHz shown in Fig.~\ref{fig:ibm-inj}(a).
The accuracy of these fits show that \ac{SchWARMA}
models can be used to effectively predict the expected survival probabilities.
This capability conveys that \ac{SchWARMA} can be a powerful tool for
parametric modeling of temporally correlated noise environments and therefore,
improved quantum system dynamics simulation informed by experimental data.

Given a \ac{FTTPS} probe sequence $k$ (with $n_k$ $R_x(\pi)$-pulses) and filter
function $g_k$, we predict the survival probability $p_k$ using the model
\begin{equation}
    p_k = \frac{1}{2}+\frac{1}{2}\exp[-g_k
    (S_{\text{nat}}(\alpha)+S_{\text{inj}})-c_1n_k-c_2n_k^2]
    \label{eq:fit-model}
\end{equation}
where $S_{\text{nat}}(\alpha)$ is a parametric model of the native dephasing noise
power spectrum (see below), $S_{\text{inj}}$ is the (known) injected spectrum, and $c_1$ and $c_2$
capture the effects of stochastic (i.e., $X-$dephasing) and coherent errors on
the $R_x(\pi)$-pulses, respectively. The pulse error terms are meant to capture the artificial high frequency features observed in the spectral estimates shown above, most notably in the case of the IBMQE results where \ac{FTTPS} without pulse-error compensation are used. We use Eq.~\eqref{eq:fit-model} to set up a
nonlinear least-squares problem to estimate the parameters $\alpha$ and $c_i$
using a set of \ac{FTTPS}. We further expand upon the motivation for the model and present additional results in the supplement.

The native noise $S_{\text{nat}}$ is, in general, qubit and system dependent.
For the APL system (Fig.~\ref{fig:schwarma_preds}~top), we found that a
combination of a Lorentzian spectrum $A/(1+\omega^2/\omega_c^2)$, with unknown
amplitude $A$ and cutoff frequency $\omega_c$, and a white noise floor with
power $\sigma^2$ captured the observed native dephasing noise (i.e., $\alpha =
\{A,\omega_c^2,\sigma^2\}$), which exhibited low-frequency energy that was not
well modeled by the white noise floor nor the control noise terms $c_i$.
For the IBM system (Fig.~\ref{fig:schwarma_preds}~bottom), we found that there
were isolated  resonances at the 0th and 8th sequences, and beyond that
the noise was well approximated by a white noise floor.  Thus, instead of
introducing additional model terms in $S_{nat}$ to account for these two
sequences (i.e., the native noise in those frequency bands), we instead ignored
these sequences in the estimation (as the introduction of two more ancillary
terms would perfectly match these points without changing the estimates for the
unknown white noise floor power $\sigma^2$ and $c_i$).  Again, we emphasize
that Fig.~\ref{fig:schwarma_preds} shows strong agreement of the
least-squares fits to the measured experimental data, and that we are only
fitting over the ancillary background and control noise terms, not on any terms
associated with the injected spectrum.\\

\noindent\textbf{Discussion}\\
Although quantum systems are commonly subject to temporally correlated noise, the impact of such noise processes on the performance of quantum algorithms is not well understood. Here, we present a novel method for engineering and injecting temporally correlated dephasing noise that can be used to elucidate key features of noise resilience and vulnerability in quantum algorithms for noise characterization, sensing, and computing. The approach enables noise injection of arbitrary noise spectra via \ac{SchWARMA}, a statistical tool for modeling, estimating, and generating semiclassical noise in quantum systems. \ac{SchWARMA} is used to synthesize engineered phase noise spectra that are introduced by adjusting the control master clock to emulate classical temporally correlated dephasing noise processes. We show that the experimental system dynamics generated by engineered noise can be predicted by \ac{SchWARMA}-based models; therefore, closing the loop between generation and injection processes. Our results indicate that \ac{SchWARMA} is a highly flexible tool for injecting engineered noise and developing experimental data-informed parametric noise models for simulating quantum system dynamics. The techniques presented here can be readily extended to multi-axis noise, non-Gaussian and nonstationary noise, and multi-qubit scenarios. This points towards the generalizability and applicability of \ac{SchWARMA} as a tool for understanding, evaluating, and improving quantum algorithms subjected to a wide range of temporally correlated noise processes relevant to current and future hardware platforms.\\

\noindent\textbf{Acknowledgements}\\
The authors acknowledge JHU/APL for the infrastructure investments required to perform these experiments. In addition, the authors acknowledge Ben Palmer, Neda Foroozani, and Kevin Osborn from the Laboratory for Physical Sciences for technical support and guidance. The authors acknowledge Russell Lake, Xian Wu, and Hsiang-Sheng Ku for chip design and fabrication. G.Q. and L.T. acknowledge funding from the DOE Office of Science, Office of Advanced Scientific Computing Research (ASCR) QCATS program, under fieldwork proposal number ERKJ347. G.Q., K.S., and L.T. acknowledge support from ARO MURI grant W911NF-18-1-0218. B.D.C. acknowledges support from DOE Office of Science ARQC Grant DE-SC0020316. J.L and D.P acknowledge support from National Science Foundation (Award No.  1839136) and the DOE through FNAL. A.M., J.E., G.Q., K.S., K.M., C.T., B.T.S., J.A.H., B.D.C., and T. S. acknowledge support from the United States Department of Defense. The views and conclusions contained in this document are those of the authors and should not be interpreted as representing the official policies, either expressly or implied, of the United States Department of Defense or the U.S. Government. 
\\

\noindent\textbf{Author Contributions}\\
K.S., B.D.C., J.A.H., and T.S. conceived of the experiments performed at APL. A.M. and J.E. performed the in-house experiments. J.L designed the device for APL experiments. B.T.S., K.M., and T.S. provided experimental assistance. G.Q., K.S., and B.D.C. conceived the IBMQE cloud-based experiments. G.Q. and L.T. performed the IBMQE experiments. K.S. designed and implemented SchWARMA prediction analysis and numerical simulations. K.S. designed the noise spectroscopy sequences with feedback from J.E. and G.Q.. A.M., J.E., G.Q., K.S., and L.T. analyzed the data. G.Q. coordinated manuscript writing with substantial contributions from A.M., J.E., K.S., K.M., C.T., and T.S.. Various aspects of the project were managed by G.Q., J.A.H., D.P., and T.S. Finally, all authors interpreted data and contributed to editing of the manuscript.


\begin{thebibliography}{10}
\expandafter\ifx\csname url\endcsname\relax
  \def\url#1{\texttt{#1}}\fi
\expandafter\ifx\csname urlprefix\endcsname\relax\def\urlprefix{URL }\fi
\providecommand{\bibinfo}[2]{#2}
\providecommand{\eprint}[2][]{\url{#2}}

\bibitem{romach2015:qns}
\bibinfo{author}{Romach, Y.} \emph{et~al.}
\newblock \bibinfo{title}{Spectroscopy of surface-induced noise using shallow
  spins in diamond}.
\newblock \emph{\bibinfo{journal}{Phys. Rev. Lett.}}
  \textbf{\bibinfo{volume}{114}}, \bibinfo{pages}{017601}
  (\bibinfo{year}{2015}).

\bibitem{Frey2017:qns}
\bibinfo{author}{Frey, V.~M.} \emph{et~al.}
\newblock \bibinfo{title}{Application of optimal band-limited control protocols
  to quantum noise sensing}.
\newblock \emph{\bibinfo{journal}{Nature Communications}}
  \textbf{\bibinfo{volume}{8}}, \bibinfo{pages}{2189} (\bibinfo{year}{2017}).

\bibitem{chan2018:qns}
\bibinfo{author}{Chan, K.~W.} \emph{et~al.}
\newblock \bibinfo{title}{Assessment of a silicon quantum dot spin qubit
  environment via noise spectroscopy}.
\newblock \emph{\bibinfo{journal}{Phys. Rev. Applied}}
  \textbf{\bibinfo{volume}{10}}, \bibinfo{pages}{044017}
  (\bibinfo{year}{2018}).

\bibitem{sung2019:qns}
\bibinfo{author}{Sung, Y.} \emph{et~al.}
\newblock \bibinfo{title}{Non-gaussian noise spectroscopy with a
  superconducting qubit sensor}.
\newblock \emph{\bibinfo{journal}{Nature Communications}}
  \textbf{\bibinfo{volume}{10}}, \bibinfo{pages}{3715} (\bibinfo{year}{2019}).

\bibitem{lupke2019:qns}
\bibinfo{author}{L{\"u}pke, U.~{\relax von}.} \emph{et~al.}
\newblock \bibinfo{title}{Two-qubit spectroscopy of spatiotemporally correlated
  quantim noise in superconducting qubits}.
\newblock \emph{\bibinfo{journal}{arXiv preprint arXiv:1912.04982}}
  (\bibinfo{year}{2019}).

\bibitem{blumekohout2011:gst}
\bibinfo{author}{Blume-Kohout, R.}, \bibinfo{author}{Gamble, J.},
  \bibinfo{author}{Nielsen, E.} \& \bibinfo{author}{et~al.}
\newblock \bibinfo{title}{Demonstration of qubit operations below a rigorous
  fault tolerance threshold with gate set tomography.}
\newblock \emph{\bibinfo{journal}{Nat Commun}} \textbf{\bibinfo{volume}{8}},
  \bibinfo{pages}{14485} (\bibinfo{year}{2017}).

\bibitem{knill2008:rb}
\bibinfo{author}{Knill, E.} \emph{et~al.}
\newblock \bibinfo{title}{Randomized benchmarking of quantum gates}.
\newblock \emph{\bibinfo{journal}{Phys. Rev. A}} \textbf{\bibinfo{volume}{77}},
  \bibinfo{pages}{012307} (\bibinfo{year}{2008}).

\bibitem{magesan2011:rb}
\bibinfo{author}{Magesan, E.}, \bibinfo{author}{Gambetta, J.~M.} \&
  \bibinfo{author}{Emerson, J.}
\newblock \bibinfo{title}{Scalable and robust randomized benchmarking of
  quantum processes}.
\newblock \emph{\bibinfo{journal}{Phys. Rev. Lett.}}
  \textbf{\bibinfo{volume}{106}}, \bibinfo{pages}{180504}
  (\bibinfo{year}{2011}).

\bibitem{lidar2014:dfs-rev}
\bibinfo{author}{Lidar, D.}
\newblock \bibinfo{title}{Review of decoherence free subspaces, noiseless
  subsystems, and dynamical decoupling}.
\newblock \emph{\bibinfo{journal}{Adv. Chem. Phys.}}
  \textbf{\bibinfo{volume}{154}}, \bibinfo{pages}{295--354}
  (\bibinfo{year}{2014}).

\bibitem{viola1998:dd}
\bibinfo{author}{Viola, L.} \& \bibinfo{author}{Lloyd, S.}
\newblock \bibinfo{title}{Dynamical suppression of decoherence in two-state
  quantum systems}.
\newblock \emph{\bibinfo{journal}{Phys. Rev. A}} \textbf{\bibinfo{volume}{58}},
  \bibinfo{pages}{2733--2744} (\bibinfo{year}{1998}).

\bibitem{viola1999:dd}
\bibinfo{author}{Viola, L.}, \bibinfo{author}{Knill, E.} \&
  \bibinfo{author}{Lloyd, S.}
\newblock \bibinfo{title}{Dynamical decoupling of open quantum systems}.
\newblock \emph{\bibinfo{journal}{Phys. Rev. Lett.}}
  \textbf{\bibinfo{volume}{82}}, \bibinfo{pages}{2417--2421}
  (\bibinfo{year}{1999}).

\bibitem{zanardi1999:dd}
\bibinfo{author}{Zanardi, P.}
\newblock \bibinfo{title}{Symmetrizing evolutions}.
\newblock \emph{\bibinfo{journal}{Physics Letters A}}
  \textbf{\bibinfo{volume}{258}}, \bibinfo{pages}{77 -- 82}
  (\bibinfo{year}{1999}).

\bibitem{viola2013:dd-rev}
\bibinfo{author}{Viola, L.}
\newblock \emph{\bibinfo{title}{Introduction to quantum dynamical decoupling}},
  \bibinfo{pages}{105–125} (\bibinfo{publisher}{Cambridge University Press},
  \bibinfo{year}{2013}).

\bibitem{shor1995:qec}
\bibinfo{author}{Shor, P.~W.}
\newblock \bibinfo{title}{Scheme for reducing decoherence in quantum computer
  memory}.
\newblock \emph{\bibinfo{journal}{Phys. Rev. A}} \textbf{\bibinfo{volume}{52}},
  \bibinfo{pages}{R2493--R2496} (\bibinfo{year}{1995}).

\bibitem{steane1996:qec}
\bibinfo{author}{Steane, A.~M.}
\newblock \bibinfo{title}{Error correcting codes in quantum theory}.
\newblock \emph{\bibinfo{journal}{Phys. Rev. Lett.}}
  \textbf{\bibinfo{volume}{77}}, \bibinfo{pages}{793--797}
  (\bibinfo{year}{1996}).

\bibitem{aliferis2006:qec}
\bibinfo{author}{Aliferis, P.}, \bibinfo{author}{Gottesman, D.} \&
  \bibinfo{author}{Preskill, J.}
\newblock \bibinfo{title}{Quantum accuracy threshold for concatenated
  distance-3 codes}.
\newblock \emph{\bibinfo{journal}{Quantum Info. Comput.}}
  \textbf{\bibinfo{volume}{6}}, \bibinfo{pages}{97–165}
  (\bibinfo{year}{2006}).

\bibitem{terhal2015:2015}
\bibinfo{author}{Terhal, B.~M.}
\newblock \bibinfo{title}{Quantum error correction for quantum memories}.
\newblock \emph{\bibinfo{journal}{Rev. Mod. Phys.}}
  \textbf{\bibinfo{volume}{87}}, \bibinfo{pages}{307--346}
  (\bibinfo{year}{2015}).

\bibitem{nielsen2011:qc}
\bibinfo{author}{Nielsen, M.~A.} \& \bibinfo{author}{Chuang, I.~L.}
\newblock \emph{\bibinfo{title}{Quantum Computation and Quantum Information:
  10th Anniversary Edition}} (\bibinfo{publisher}{Cambridge University Press},
  \bibinfo{address}{USA}, \bibinfo{year}{2011}), \bibinfo{edition}{10th} edn.

\bibitem{chuang1995:qc-err}
\bibinfo{author}{Chuang, I.~L.}, \bibinfo{author}{Laflamme, R.},
  \bibinfo{author}{Shor, P.~W.} \& \bibinfo{author}{Zurek, W.~H.}
\newblock \bibinfo{title}{Quantum computers, factoring, and decoherence}.
\newblock \emph{\bibinfo{journal}{Science}} \textbf{\bibinfo{volume}{270}},
  \bibinfo{pages}{1633--1635} (\bibinfo{year}{1995}).
\newblock
  \eprint{https://science.sciencemag.org/content/270/5242/1633.full.pdf}.

\bibitem{knill1998:qc-err}
\bibinfo{author}{Knill, E.}, \bibinfo{author}{Laflamme, R.} \&
  \bibinfo{author}{Zurek, W.~H.}
\newblock \bibinfo{title}{Resilient quantum computation: error models and
  thresholds}.
\newblock \emph{\bibinfo{journal}{Proc. R. Soc. Lond.}}
  \textbf{\bibinfo{volume}{454}}, \bibinfo{pages}{365–384}
  (\bibinfo{year}{1998}).

\bibitem{Knill1998:qc-err2}
\bibinfo{author}{Knill, E.}, \bibinfo{author}{Laflamme, R.} \&
  \bibinfo{author}{Zurek, W.~H.}
\newblock \bibinfo{title}{Resilient quantum computation}.
\newblock \emph{\bibinfo{journal}{Science}} \textbf{\bibinfo{volume}{279}},
  \bibinfo{pages}{342--345} (\bibinfo{year}{1998}).
\newblock
  \eprint{https://science.sciencemag.org/content/279/5349/342.full.pdf}.

\bibitem{colless2018:qc-err}
\bibinfo{author}{Colless, J.~I.} \emph{et~al.}
\newblock \bibinfo{title}{Computation of molecular spectra on a quantum
  processor with an error-resilient algorithm}.
\newblock \emph{\bibinfo{journal}{Phys. Rev. X}} \textbf{\bibinfo{volume}{8}},
  \bibinfo{pages}{011021} (\bibinfo{year}{2018}).

\bibitem{mcclean2017:qc-err}
\bibinfo{author}{McClean, J.~R.}, \bibinfo{author}{Kimchi-Schwartz, M.~E.},
  \bibinfo{author}{Carter, J.} \& \bibinfo{author}{de~Jong, W.~A.}
\newblock \bibinfo{title}{Hybrid quantum-classical hierarchy for mitigation of
  decoherence and determination of excited states}.
\newblock \emph{\bibinfo{journal}{Phys. Rev. A}} \textbf{\bibinfo{volume}{95}},
  \bibinfo{pages}{042308} (\bibinfo{year}{2017}).

\bibitem{mavadia2018:gst-err}
\bibinfo{author}{Mavadia, S.}, \bibinfo{author}{Edmunds, C.},
  \bibinfo{author}{Hempel, C.} \& \bibinfo{author}{et~al.}
\newblock \bibinfo{title}{Experimental quantum verification in the presence of
  temporally correlated noise}.
\newblock \emph{\bibinfo{journal}{NPJ Quant. Inf.}}
  \textbf{\bibinfo{volume}{4}} (\bibinfo{year}{2018}).

\bibitem{soare2014:eng-noise}
\bibinfo{author}{Soare, A.} \emph{et~al.}
\newblock \bibinfo{title}{Experimental bath engineering for quantitative
  studies of quantum control}.
\newblock \emph{\bibinfo{journal}{Physical Review A}}
  \textbf{\bibinfo{volume}{8}}, \bibinfo{pages}{042329} (\bibinfo{year}{2014}).

\bibitem{schwarma}
\bibinfo{author}{Schultz, K.}, \bibinfo{author}{Quiroz, G.},
  \bibinfo{author}{Titum, P.} \& \bibinfo{author}{Clader, B.~D.}
\newblock \bibinfo{title}{{SchWARMA}: A model-based approach for
  time-correlated noise in quantum circuits}.
\newblock \emph{\bibinfo{journal}{arXiv preprint arXiv:2010.04580}}
  (\bibinfo{year}{2020}).

\bibitem{ibmqe}
\bibinfo{title}{{IBM Quantum Experience}}.
\newblock \bibinfo{howpublished}{{https://quantum-computing.ibm.com/}}.

\bibitem{alvarez2011:qns}
\bibinfo{author}{\'Alvarez, G.~A.} \& \bibinfo{author}{Suter, D.}
\newblock \bibinfo{title}{Measuring the spectrum of colored noise by dynamical
  decoupling}.
\newblock \emph{\bibinfo{journal}{Phys. Rev. Lett.}}
  \textbf{\bibinfo{volume}{107}}, \bibinfo{pages}{230501}
  (\bibinfo{year}{2011}).

\bibitem{ball2016:clock}
\bibinfo{author}{Ball, H.}, \bibinfo{author}{Oliver, W.} \&
  \bibinfo{author}{Biercuk, M.}
\newblock \bibinfo{title}{The role of master clock stability in quantum
  information processing}.
\newblock \emph{\bibinfo{journal}{NPJ Quantum Inf}}
  \textbf{\bibinfo{volume}{2}}, \bibinfo{pages}{16033} (\bibinfo{year}{2016}).

\bibitem{mckay2017:gates}
\bibinfo{author}{McKay, D.~C.}, \bibinfo{author}{Wood, C.~J.},
  \bibinfo{author}{Sheldon, S.}, \bibinfo{author}{Chow, J.~M.} \&
  \bibinfo{author}{Gambetta, J.~M.}
\newblock \bibinfo{title}{Efficient $z$ gates for quantum computing}.
\newblock \emph{\bibinfo{journal}{Phys. Rev. A}} \textbf{\bibinfo{volume}{96}},
  \bibinfo{pages}{022330} (\bibinfo{year}{2017}).

\bibitem{burnett2014:fnoise}
\bibinfo{author}{Meeson, J.}, \bibinfo{author}{Tzalenchuk, A.~{\relax Ya}.} \&
  \bibinfo{author}{Lindstr{\"{o}}m, T.}
\newblock \bibinfo{title}{Evidence for interacting two-level systems from the
  1/f noise of a superconducting resonator}.
\newblock \emph{\bibinfo{journal}{Nature Communications}}
  \textbf{\bibinfo{volume}{5}}, \bibinfo{pages}{4119} (\bibinfo{year}{2014}).

\bibitem{burnett2019:fnoise}
\bibinfo{author}{Burnett, J.~J.} \emph{et~al.}
\newblock \bibinfo{title}{Decoherence benchmarking of superconducting qubits}.
\newblock \emph{\bibinfo{journal}{npj Quantum Information}}
  \textbf{\bibinfo{volume}{5}}, \bibinfo{pages}{54} (\bibinfo{year}{2019}).

\bibitem{muller2015:fnoise}
\bibinfo{author}{M{\"u}ller, C.}, \bibinfo{author}{Lisenfeld, J.},
  \bibinfo{author}{Shnirman, A.} \& \bibinfo{author}{S., P.}
\newblock \bibinfo{title}{Non-gaussian noise spectroscopy with a
  superconducting qubit sensor}.
\newblock \emph{\bibinfo{journal}{Physical Review B}}
  \textbf{\bibinfo{volume}{92}}, \bibinfo{pages}{035442}
  (\bibinfo{year}{2015}).

\bibitem{bylander2011:fnoise}
\bibinfo{author}{Bylander, J.} \emph{et~al.}
\newblock \bibinfo{title}{Noise spectroscopy through dynamical decoupling with
  a superconducting flux qubit}.
\newblock \emph{\bibinfo{journal}{Nature Physics}}
  \textbf{\bibinfo{volume}{7}}, \bibinfo{pages}{656} (\bibinfo{year}{2019}).

\bibitem{yan2013:fnoise}
\bibinfo{author}{Yan, F.} \emph{et~al.}
\newblock \bibinfo{title}{Rotationg-frame relaxation as a noise spectrum
  analyser of a superconducting qubit undergoing driven evolution}.
\newblock \emph{\bibinfo{journal}{Nature Communications}}
  \textbf{\bibinfo{volume}{4}}, \bibinfo{pages}{22337} (\bibinfo{year}{2013}).

\bibitem{basset2014:fnoise}
\bibinfo{author}{Basset, J.} \emph{et~al.}
\newblock \bibinfo{title}{Evaluating charge noise acting on semiconductor
  quantum dots in the circuit quantum electrodynamics architecture}.
\newblock \emph{\bibinfo{journal}{Applied Physics Letters}}
  \textbf{\bibinfo{volume}{105}}, \bibinfo{pages}{063105}
  (\bibinfo{year}{2014}).

\bibitem{struck2020:fnoise}
\bibinfo{author}{Struck, T.} \emph{et~al.}
\newblock \bibinfo{title}{Low-frequency spin qubit energy splitting noise in
  highly purified $^{28}$si/sige}.
\newblock \emph{\bibinfo{journal}{npj Quantum Information}}
  \textbf{\bibinfo{volume}{6}}, \bibinfo{pages}{40} (\bibinfo{year}{2020}).

\bibitem{yan2016:qns}
\bibinfo{author}{Yan, F.}, \bibinfo{author}{Gustavsson, S.},
  \bibinfo{author}{Kamal, A.} \& \bibinfo{author}{et~al.}
\newblock \bibinfo{title}{The flux qubit revisited to enhance coherence and
  reproducibility}.
\newblock \emph{\bibinfo{journal}{Nat Commun}}  (\bibinfo{year}{2016}).

\bibitem{thomas2020:pulses}
\bibinfo{author}{Alexander, T.~A.} \emph{et~al.}
\newblock \bibinfo{title}{Qiskit-pulse: programming quantum computers through
  the cloud with pulses}.
\newblock \emph{\bibinfo{journal}{Quantum Science and Technology}}
  (\bibinfo{year}{2020}).

\bibitem{vold1973:comp}
\bibinfo{author}{Vold, R.~L.} \& \bibinfo{author}{Simon, H.~E.}
\newblock \bibinfo{title}{Errors in measurements of transverse relaxation
  rates}.
\newblock \emph{\bibinfo{journal}{Journal of Magnetic Resonance (1969)}}
  \textbf{\bibinfo{volume}{11}}, \bibinfo{pages}{283 -- 298}
  (\bibinfo{year}{1973}).

\bibitem{levitt1981:comp}
\bibinfo{author}{Levitt, M.~H.} \& \bibinfo{author}{Freeman, R.}
\newblock \bibinfo{title}{Composite pulse decoupling}.
\newblock \emph{\bibinfo{journal}{Journal of Magnetic Resonance (1969)}}
  \textbf{\bibinfo{volume}{43}}, \bibinfo{pages}{502 -- 507}
  (\bibinfo{year}{1981}).

\bibitem{taylor2007:thesis}
\bibinfo{author}{Smith, S.~T.}
\newblock \emph{\bibinfo{title}{Bounded-strength dynamical control of a qubit
  based on Eulerian cycles}}.
\newblock Ph.D. thesis (\bibinfo{year}{2007}).

\bibitem{quiroz2013:gadd}
\bibinfo{author}{Quiroz, G.} \& \bibinfo{author}{Lidar, D.~A.}
\newblock \bibinfo{title}{Optimized dynamical decoupling via genetic
  algorithms}.
\newblock \emph{\bibinfo{journal}{Phys. Rev. A}} \textbf{\bibinfo{volume}{88}},
  \bibinfo{pages}{052306} (\bibinfo{year}{2013}).

\end{thebibliography}
\end{document}